%% file: VQS-Merged.tex
\documentclass[aps,prl,twocolumn,superscriptaddress]{revtex4-1}
\usepackage[T1]{fontenc}
\usepackage[latin9]{inputenc}
\usepackage[english]{babel} 

\usepackage{amsmath}
\usepackage{verbatim}
\usepackage{graphicx}
\usepackage{color}
\usepackage{xcolor}
\usepackage{setspace}
\usepackage{braket}
\usepackage{bbold}
\usepackage{hyperref}

\usepackage{tipa}
\usepackage{amssymb}
\usepackage{wasysym}
\usepackage{bbold}
\usepackage[abs]{overpic}
\usepackage{cancel}
\graphicspath{{../}}

\makeatletter

 \definecolor{boxback}{HTML}{FFF8B5}
 
 \definecolor{applegreen}{rgb}{0, 0.5, 0.0}

\newcommand{\ict}{\affiliation{Center for Quantum Physics, University of Innsbruck, Austria}}
\newcommand{\iqoqi}{\affiliation{Institute for Quantum Optics and Quantum Information of the Austrian Academy of Sciences,  Innsbruck, Austria}}
\newcommand{\jila}{\affiliation{JILA, University of Colorado and National Institute of Standards and Technology, and Department of Physics, University of Colorado, Boulder,  CO 80309, USA}}
\newcommand{\ctqm}{\affiliation{Center for Theory of Quantum Matter, University of Colorado, Boulder, CO 80309, USA}}

\definecolor{smoothred}{HTML}{C5232F}
\definecolor{mygreen}{rgb}{0,0.5,0}
\definecolor{myblue}{rgb}{0,0,0.75}
\definecolor{mymagenta}{cmyk}{0,1,0,0.12}

\newcommand{\paramvec}{\boldsymbol{\theta}}

\DeclareMathOperator{\Tr}{Tr}

\begin{document}
\title{Variational spin-squeezing algorithms on programmable quantum sensors}

\author{Raphael Kaubruegger}\ict \iqoqi
\author{Pietro Silvi}\ict \iqoqi
\author{Christian Kokail}\ict \iqoqi
\author{Rick van Bijnen}\ict \iqoqi
\author{Ana Maria Rey}\jila \ctqm
\author{Jun Ye}\jila
\author{Adam M. Kaufman}\jila
\author{Peter Zoller}\ict \iqoqi

\date{\today}

\begin{abstract}
Arrays of atoms trapped in optical tweezers combine features of programmable analog quantum simulators with atomic quantum sensors. Here we propose variational quantum algorithms, tailored for tweezer arrays as programmable quantum sensors, capable of generating entangled states on-demand for precision metrology. The scheme is designed to generate metrological enhancement by optimizing it in a feedback loop on the quantum device itself, thus preparing the best entangled states given the available quantum resources. We apply our ideas to generate spin-squeezed states on Sr atom tweezer arrays, where finite-range
interactions are generated through Rydberg dressing.
The complexity of experimental variational optimization of our quantum circuits is expected to scale favorably with system size.
We numerically show our approach to be robust to noise, and surpassing known protocols.
\end{abstract}

\maketitle

Optical tweezer arrays of neutral atoms provide a bottom-up approach to assemble and design quantum many-body systems `atom by atom'. The flexibility and universality of tweezers, as a novel tool to engineer atomic and molecular quantum devices, is demonstrated by recent experiments, which range from realization of `programmable' analog quantum simulators for spin-models in tweezer arrays \cite{BrowaeysQSim2016, LukinQSim2017, NiDoylePolar2019}, to first demonstrations of potential tweezer-based clocks \cite{Cooper2018, Norcia2018, Saskin2019, Covey2019, Norcia2019, Madjarov2019}. Atomic many-body systems designed around tweezer platforms thus offer the unique possibility of combining, on the same physical device, programablility to generate many-particle entangled states, and adopting these states as a quantum resource in precision measurement, exhibiting quantum advantage provided by entanglement. With near future experiments promising a scaling to hundreds of atoms, the challenge is to design and run quantum algorithms, which efficiently generate entangled states of interest for precision measurements, given the -- in general non-universal -- entangling resources available on  `programmable' quantum sensors, and Rydberg tweezer arrays in particular.

\begin{figure}[t!]
\includegraphics[width=\columnwidth]{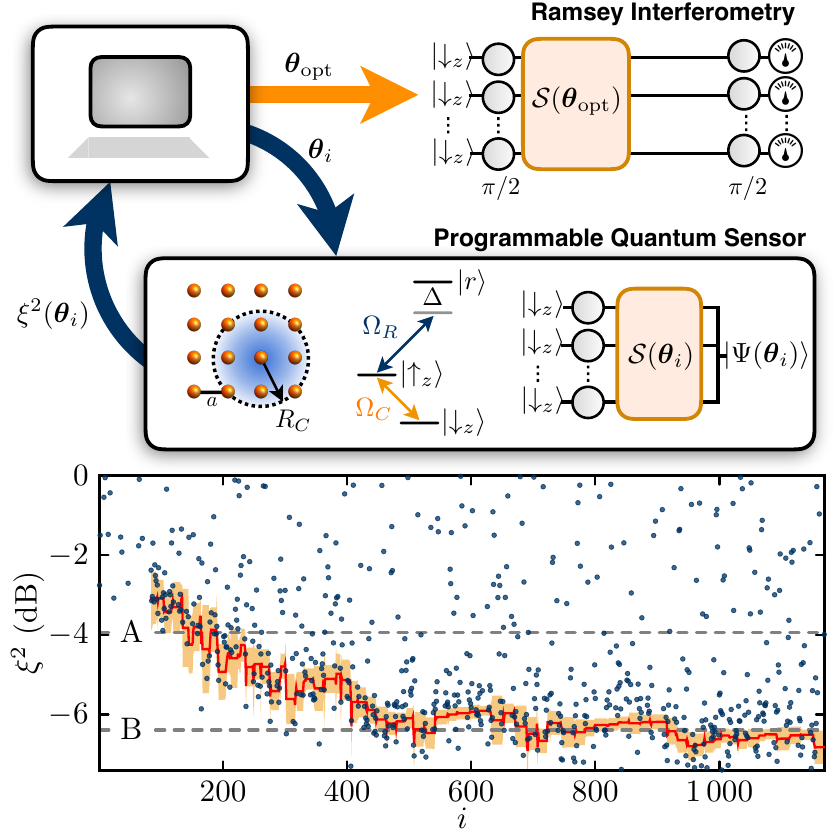}
\caption{\label{fig:one}
Hybrid classical-quantum optimization on a programmable quantum sensor prepares spin-squeezed states for Ramsey interferometry. Variational control parameters $\boldsymbol{\theta}$  generate trial states $\ket{\Psi(\paramvec)}=\mathcal{S}(\paramvec)R_y(\pi/2)\ket{\downarrow_z}^{\otimes N}$.  The squeezing parameter $\xi(\boldsymbol{\theta})$  serves as cost function for  classical optimization.
Bottom: Optimization run on a $4\times 4$ array with $R_C/a=1.5$, using $\approx10^5$ simulated experimental runs, showing measured spin-squeezing of the trial states (blue dots, one iteration = 100 measurements).  A and B indicate theoretically obtainable squeezing with finite- and infinite-range one axis twisting \cite{KitagawaSpinsqueeze1993, PohlSqueeze2014}. Red line and orange error band represent the predicted squeezing minima at iteration $i$ of the search algorithm. 
}
\end{figure}

Here we propose hybrid classical-quantum algorithms \cite{VQARahmani2013,Peruzzo2014,VQATroyer2015,VQAMcClean2016,Kandala2017,VQSError2017,MollGambetta2018,Kokail2019,Klco_2018,QAOAPichler2018} to be run as a quantum feedback loop to generate the \textit{best} entangled states for the given platform, yielding precision enhancement beyond the standard quantum limit (SQL) \cite{Giovannetti2004, RMPSqueezing}. 
The variational many-body wavefunction is optimized on the quantum device itself, in terms of a relevant cost function quantifying the metrological enhancement.
Performing the optimization on the physical platform will yield the best entangled state achievable in the presence of the actual imperfections and noise, thus outperforming optimization loops purely based on numerical simulations \cite{OMalley2016}.
Moreover, since near-term quantum devices are expected to soon operate in regimes beyond the reach of numerical simulations, the optimization loop can ultimately \textit{only} be run directly on the programmable quantum sensor. 
After optimization, the resulting optimal wavefunction can be re-prepared on demand on the quantum sensor
\footnote{
Performing occasional re-optimization steps, via local search algorithm starting from the previous optimal solution, can account for low frequency noise, fluctuating at timescales longer than the optimization runtime (slow drifts in the experimental quantities).},
directly available for high precision measurement (see Fig.~\ref{fig:one}).
In this work we specifically target the optimized preparation of spin-squeezed states \cite{WinelandSpinsqueeze1992, KitagawaSpinsqueeze1993, RMPSqueezing}, a class of entangled states enhancing measurement precision of atomic Ramsey interferometers, including programmable tweezer clocks \cite{Norcia2019, Madjarov2019}. 
{\it Ramsey interferometry --}
The Ramsey sequence, acting on a two-level atom, described by spin-1/2 states $\{\ket{\downarrow_z},\ket{\uparrow_z}\}$, and corresponding spin operators $\boldsymbol{s} \equiv \hbar (\sigma_x,\sigma_y,\sigma_z)/2$, starts with an initial $\pi/2$-pulse that creates a coherent superposition $(\ket{\downarrow_z}+\ket{\uparrow_z})/\sqrt{2}=\ket{\uparrow_x}$. In the subsequent interrogation time, $\ket{\downarrow_z}$ and $\ket{\uparrow_z}$ acquire a relative phase $\varphi$, encoding the quantity to be measured. The final $\pi/2$-pulse transfers this phase difference into a measurable state population difference.
In this context, spin-squeezed states are a well-known family of entangled states enhancing the phase sensitivity 
over $N$ uncorrelated atoms \cite{WinelandSpinsqueeze1992}. Here, we prepare spin-squeezed states via an entangling squeezing operation $\mathcal{S}(\boldsymbol{\theta})$ realizable on the programmable quantum sensor
(see Fig.~\ref{fig:one} ${\bf a}$). The variational quantum algorithm optimizes the classical control variables $\boldsymbol{\theta}$, parametrizing
$\mathcal{S}(\boldsymbol{\theta})$, to achieve optimal spin-squeezing.

{\it Spin squeezing --}
The achievable phase sensitivity $\Delta \varphi = \xi/\sqrt{N}$, is quantified by the spin squeezing parameter \cite{WinelandSpinsqueeze1992}
\begin{align}
    \xi^2(\boldsymbol{\theta})=N\frac{\left(\Delta J_{\perp, \min}\right)_{\boldsymbol{\theta}}^2}{\left|\braket{\boldsymbol{J}}\right|_{\boldsymbol{\theta}}^2},
    \label{eq:SqueezingParameter}
\end{align}
to be minimized on the quantum sensor. Here $\boldsymbol{J}=\sum_{i=1}^N\boldsymbol{s}_i$ denotes the collective spin vector 
associated with an ensemble of $N$ two-level atoms and $\left(\Delta J_{\perp, \min}\right)^2_{\boldsymbol{\theta}}\equiv\langle J_{\perp, \min}^2\rangle_{\boldsymbol{\theta}}-\langle J_{\perp, \min}\rangle^2_{\boldsymbol{\theta}}$ quantifies the minimal spin fluctuation orthogonal to the Bloch vector $\braket{\boldsymbol{J}}_{\boldsymbol{\theta}}$.
The expectation values in \eqref{eq:SqueezingParameter} are to be estimated, with respect to the variational wavefunction $|\Psi(\boldsymbol{\theta})\rangle = \mathcal{S}(\boldsymbol{\theta}) \ket{\uparrow_x}^{\otimes N}$.
In atom-tweezer arrays, $\langle J_{\perp, \min}^2\rangle_{\boldsymbol{\theta}}$ is assembled from two-body correlation functions, directly measurable due to single site resolution.

Mechanisms known to generate spin squeezing are one-axis twisting $S_1(\tau) = \exp [ - i \tau J_z^2 ]$ (OAT), implemented in experiments \cite{Sackett2000, Leibfried2005, Monz2011, Bohnet2016, Esteve2008, Riedel2010, Luecke2011, Hamley2012,  Berrada2013, Zou2018, Appel2009, Leroux2010, Chen2014, Sewell2014, Barontini2015, Hosten2016}, and two-axis twisting (TAT) $S_2(\tau) = \exp [ - i \tau (J_z^2 - J_y^2) ]$ \cite{KitagawaSpinsqueeze1993}. Various theoretical studies \cite{LiuYouSqueeze2011,Shen2013,SqueezingTrotter2014,OptimalControlSqueezing,BouchouleTATRYdberg2002} employ time-dependent dynamics to generate effective TAT, and more generally spin-squeezing up to the Heisenberg limit $\xi^2 \simeq 1/N$. We note that these approaches rely on infinite-range interactions,  conserving symmetries that constrain the dynamics to the particle permutation
symmetric subspace, of linear dimension (N+1), where the maximally spin-squeezed states are known to be located \cite{Brif1996, OptimalSqueezing} (see supplemental material $-$ SM). In contrast, dynamics originating from \emph{finite-range} interactions, such as Rydberg dressing, may explore an exponentially large Hilbert space.

\begin{figure}[t]
\includegraphics[width=\columnwidth]{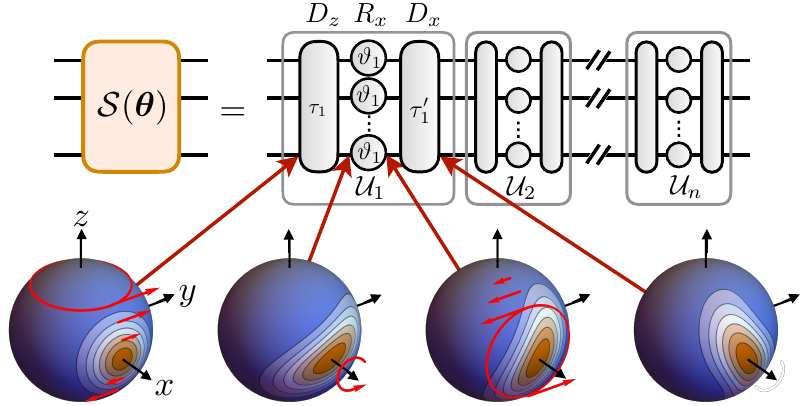}
\caption{
Quantum circuit representing the variational squeezing operation. The squeezer $\mathcal{S} (\boldsymbol{\theta})$ is a sequence of $n$ unitary layers ${\mathcal U} _i$ ($i=1,\ldots,n$). Each ${\mathcal U} _i$ is further decomposed into two entangling interactions $D_z$ and $D_x$ and one global rotation $R_x$, as in Eq.~\eqref{eq:layergate}, controlled by three variational parameters $\tau_i,\vartheta_i,\tau_i\prime$.
Typical Husimi distributions (see SM) of the quantum state before each gate, are displayed on a generalized Bloch sphere. The action of the gates onto the permutation invariant subspace is indicated by red arrows.
\label{fig:Two}
}
\end{figure}

{\it Programmable quantum sensor --} 
The tweezer-based Sr platform combines the advantages of a state of the art atomic clock \cite{RMPOpticlClocks, Campbell2017}, with the possibility of programming entangling operations \cite{PohlSqueeze2014}. Interactions among the atoms can be engineered via Rydberg dressing~\cite{PohlRydress2010, PupilloRydress2010, Honer2010, Rolston2010, Jau2015, Zeiher2016, Zeiher2017, Bounds2018, Arias2019}, where an off-resonant coupling $\Omega_{R}/\Delta \ll 1$ of the $\ket{\uparrow_z}$ clock level with a Rydberg state induces a distance-dependent pairwise energy shift. Here $\Omega_R$ denotes the Rabi frequency and $\Delta$ the detuning of the dressing laser.  The resulting interaction Hamiltonian is of the form $H_D = \sum_{i,j<i} V_{ij} s_i^z s_j^z+\sum_i \delta_i s_i^z$, with pairwise interaction potential
\begin{align}
V_{ij}=V_0\frac{R_C^6}{|\boldsymbol{r}_i-\boldsymbol{r}_j|^6+R_C^6},
\end{align}
between two particles at positions $\boldsymbol{r}_{i,j}$, where $V_0=\left(\Omega_R/2\Delta\right)^3\hbar \Omega_R$ and $R_C=\left|C_6/2\hbar \Delta\right|^{1/6}$ are related to the laser parameter and the $C_6$ van-der-Waals coefficient of the Rydberg state. 
The amount of spin-squeezing generated from these \emph{finite-range} interactions in a OAT protocol  has been studied in Ref.~\cite{PohlSqueeze2014}, and  we will refer to this as finite-range OAT (fOAT).

Below we design a variational circuit,  
$\mathcal{S}(\boldsymbol{\theta})=\mathcal{U}_n\dots \mathcal{U}_1$, to optimize spin squeezing from the physical resources described above. The circuit
$\mathcal{S}(\boldsymbol{\theta})=\mathcal{U}_n\dots \mathcal{U}_1$ comprises a sequence of $n$ unitary layers in which each $\mathcal{U}_i$ is composed of quantum operations of the form 
\begin{align}
    \mathcal{U}_i = D_x(\tau_i')R_x(\vartheta_i)D_z(\tau_i),
\label{eq:layergate}
\end{align}
with $\boldsymbol{\theta} = \{ \tau_1, \vartheta_1, \tau_1',\ldots,\tau_n, \vartheta_n, \tau_n' \}$. 
The fundamental building blocks of each layer are the interaction gates $D_{x,z}(\tau) = \exp [ -i \tau \sum_{i, j<i} V_{ij} s_i^{x,z} s_j^{x,z}]$ which can be obtained from  bare dressing in combination with global rotations $R_{x,y,z}(\vartheta_i) = \exp\left[-i \vartheta_i J_{x,y,z}\right]$ (see SM). The design of this circuit and its building blocks is motivated by the following requirements: (i) The sequence is assembled from global gates only. Hence, the number of variational parameters ($3n$) does not increase with the system size $N$.
(ii) The unitaries $\mathcal{U}_i$ are designed to preserve the direction of the collective spin $\langle \boldsymbol{J}\rangle_{\boldsymbol{\theta}} / |\langle \boldsymbol{J}\rangle_{\boldsymbol{\theta}}|$ to be oriented along the $x$ axis. This removes the overhead of determining the direction of the Bloch vectors  via extra measurements after each variational step (see SM). Eq.~\eqref{eq:layergate} describes the most general gate sequence satisfying these requirements.

We visualize in Fig.~\ref{fig:Two} the action of the variational gate sequence, by means of the Husimi distribution plotted on a generalized Bloch sphere. The Husimi distribution displays the overlap of the time evolved state with coherent spin states with $|\boldsymbol{J}|=N/2$ (see SM). In this representation, a good SSS according to Eq.~\eqref{eq:SqueezingParameter} forms a narrow vertical ellipse. 
The actions of the gates can now also be understood: $D_z$ shears the Husimi distribution, the $R_x$ gate performs rigid rotations, whereas $D_x$ causes a winding around the $x$-axis. Together, these transformations enable optimal spin squeezing while passing through transient non-elliptic states \cite{StrobelNonGaussian2014} (see fourth sphere in Fig.~\ref{fig:Two}).

In the following, we quantify the spin squeezing, attainable by our variational ansatz on different tweezer array geometries. First, we numerically emulate the feedback loop optimization under realistic experimental conditions, imposing a finite number of experimental runs on the quantum device. Afterwards, we provide a more detailed analysis of the performance and robustness of the variational gate sequence. 

\begin{figure}[t]
\includegraphics[]{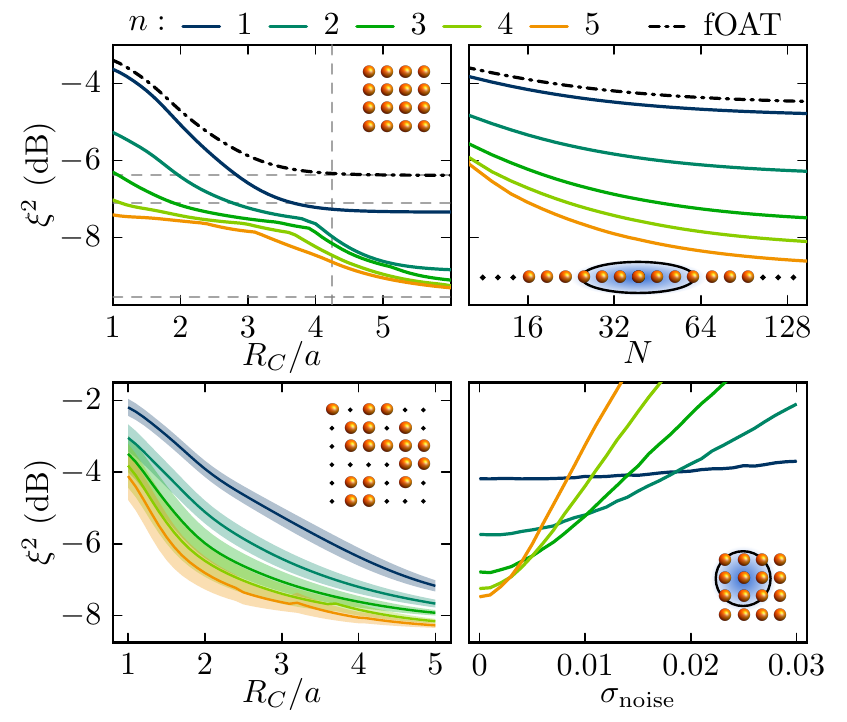}
\caption{
Exact optimization results for circuit depths from $n =$ 1 to 5 layers, compared with fOAT \protect\cite{PohlSqueeze2014} (dot dashed line).  Top left panel: optimized squeezing parameter $\xi^2$ in a $4\times4$ square array as a function of the interaction radius $R_c/a$. The vertical dashed line indicates the diagonal $R^{\star} = \sqrt{18}a$,
while the horizontal ones show, from top to bottom, the squeezing obtainable with OAT, TAT and the fundamental squeezing limit $\xi_{\lim}^2(N)=\frac{2}{N+2}$. Top right panel: optimized squeezing parameter in a 1D array, as a function of the particle number $N$, at fixed $R_C/a=3$. Bottom left panel: optimization for a random realization, displayed in the inset, of a half-filled $6 \times 6$ square array. The data points display averages over various random half-filling realizations, at the optimal pulse sequence. Bottom right panel: impact of normally-distributed control noise fluctuating with a relative standard deviation $\sigma_{\text{noise}}$ around the optimal values.
\label{fig:OptExactExps}}
\end{figure}

{\it 2D arrays: single optimization run with shot noise --}
We perform a numerical simulation of the feedback-loop optimization on a 4$\times$4 square lattice with short-ranged interactions, where we chose a circuit depth of $n=4$, corresponding to 12 variational parameters. During optimization, the cost function is estimated on the simulated experiment, as a statistical average over several runs. Each run consists of: (i) preparation of the initial state $\ket{\downarrow_z}^{\otimes N}$, (ii) coherent quantum dynamics controlled by trial parameters $\paramvec_i$ of the current iteration $i$ of the search algorithm, and (iii) quantum projective measurements, performed in parallel on every spin (all spins are measured in the same basis, either $x$ or $y$, to estimate numerator and denominator of $\xi^2(\paramvec_i)=N \langle J_y^2\rangle_{\boldsymbol{\theta}_i}/\braket{J_x}^2_{\boldsymbol{\theta}_i}$ respectively).
Fig.~\ref{fig:one} shows an optimization trajectory, employing $100$ runs for a single cost function evaluation, and restricting the total number of runs to $\sim10^5$, compatible with current repetition rates of Sr tweezer platforms of $\sim$1$\,$Hz. 
The number of runs required for a single cost function evaluation at fixed precision does not increase with $N$ (see SM).
Our analysis demonstrates that even in the presence of noisy cost function evaluations, we are able to obtain considerable spin-squeezing, surpassing the squeezing attainable from infinite-range OAT, with short-ranged interactions. The optimization algorithm that we adopt is a modified version of the DIRECT algorithm \cite{Direct1, Direct2, Direct3, Direct4}, as has been implemented in experiments on hybrid quantum-classical simulation (see methods section in Ref.~\cite{Kokail2019}).

\begin{figure}[t!]
\includegraphics[width=\columnwidth]{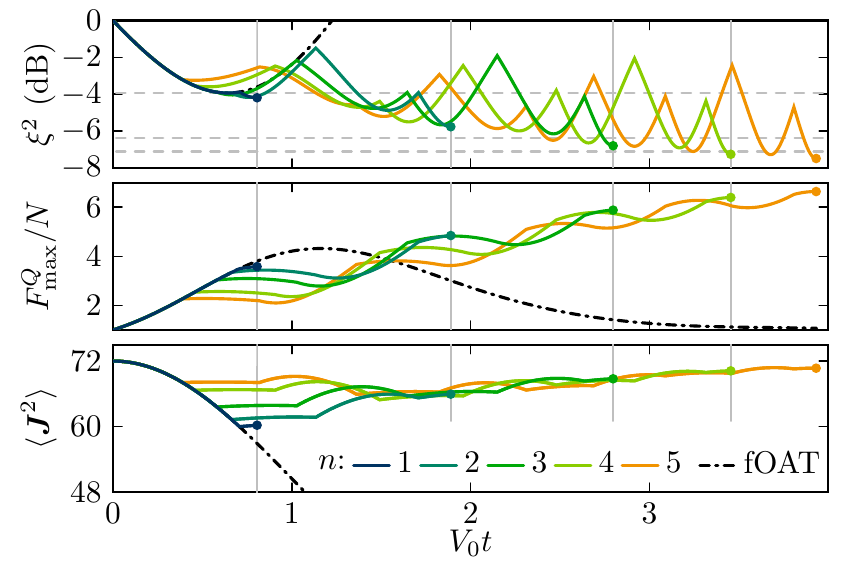}
\caption{\label{fig:RealTimeEvolution}
Real-time dynamics of optimized pulse sequences, with different squeezer depths $n$ on a $4\times4$ square array ($R_C/a=1.5$). Rotation pulses are treated as instantaneous, so that $t \in [0, \sum_{i=1}^n(\tau_i+\tau_i')]$ captures the cumulated interaction time.
Dot-dashed lines show evolution under a single $D_z$ interaction.  
The displayed quantities are the (rotationally invariant) spin squeezing $\xi^2$ (top panel), quantum Fisher information $F^Q_{\max}/N$ (middle) and total angular momentum $\langle \boldsymbol{J}^2\rangle$ (bottom).  The horizontal dashed lines in the top panel correspond to the spin squeezing obtainable with fOAT, infinite-range OAT and TAT respectively, top to bottom. Vertical lines mark total interaction time for various circuit depths.}
\end{figure}
   
{\it 2D arrays: exact results--}
We now analyze the squeezing theoretically attainable with our variational circuit, at finite circuit depth $n$ and interaction radius $R_C$.
Fig.~\ref{fig:OptExactExps} displays results from numerical optimizations on a 4x4 square lattice, where we observe
metrological enhancement over fOAT \cite{PohlSqueeze2014}.
Our analysis reveals two distinct $R_C$ regimes, roughly separated by the square array diagonal $R^{\star} = \sqrt{2}L$, with $L = 3a$ being the edge length.
While for $R_C> R^{\star}$ interactions are all-to-all and the particle permutation symmetry is approximately protected, for $R_C < R^{\star}$ the variational ansatz covers an exponentially large Hilbert space, and restoring the symmetry requires deeper circuits.
This property is clearly visible in Fig.~\ref{fig:RealTimeEvolution}, where three quantities are plotted as a function of the effective interaction time (see SM) during an optimized evolution, for $R_c/a  = 1.5$. The bottom panel tracks the total angular momentum $\langle \boldsymbol{J}^2\rangle$, which stays large and recovers values close to its maximum $\max\langle \boldsymbol{J}^2\rangle=(N/2+1)N/2$, indicating a restoration of permutation symmetry.
The top panel displays the rotationally-invariant spin squeezing parameter. We observe that the variational state passes through intermediate states with reduced squeezing, eventually surpassing the squeezing limits of fOAT, OAT and TAT for sufficiently large depths $n$.
In the middle panel we plot the  the quantum Fisher information (QFI) $ F^Q_{\max}=\max_{\boldsymbol{n}} \frac{4}{\ |\boldsymbol{n}|^2} [\langle \left(\boldsymbol{n} \cdot \boldsymbol{J} \right)^2\rangle -\langle \boldsymbol{n} \cdot \boldsymbol{J}\rangle^2 ]$, a measure of metrological enhancement \cite{CavesFisherinformation} and a witness for $k$-partite entanglement if $F^Q_{\max}/N>k$ \cite{QFI_Entanglement_1, QFI_Entanglement_2}. We observe that, compared to $\xi^2$, the QFI is a smoother function of time, a hint that the optimal evolution passes through transient non-Gaussian states (see SM).

{\it 1D arrays --} 
In order to investigate the performance of the protocol at large particle numbers, in Fig.~\ref{fig:OptExactExps} (top right panel) we show numerical simulations on 1D arrays up to 150 sites at fixed $R_C/a=3$.
Since the low connectivity of 1D systems yields slower entanglement generation, a power-law scaling of the squeezing parameter $\xi^2(N)$ with the system size is not expected \cite{PohlSqueeze2014, FiniteRangeSqueezing}. However, the 1D geometry allows the simulation of large system sizes using matrix product states (MPS) \cite{PollmannMPOExponential}.
The plot displays the optimized spin squeezing as a function of the particle number. We observe moderate improvement when more particles are added to the system. Instead, significantly more spin-squeezing is generated if the circuit depth is increased. We expect this to hold also for 2D systems. Additionally we observe that parameters, optimized for systems where the bulk dominates over the boundary, can be directly used for larger systems, delivering increased squeezing at larger particle numbers.

{\it Imperfections and decoherence --}
In a real experiment, optimal state preparation will be influenced by noise and imperfections. 
Below, we study the effect of a stochastically-loaded array, and verify that the squeezing behaves smoothly under Gaussian control noise. 

In general, tweezer arrays are loaded stochastically, and even with defect-removal protocols imperfections will persist. In Fig.~\ref{fig:OptExactExps} (bottom left panel) we consider a randomly half-filled 6$\times$6 square array, and optimize the squeezer for a specific filling pattern, shown in the inset. The optimal pulse sequence is  applied to various half-filling configurations. The data points show the average and standard deviation over filling realizations. Despite the inhomogeneous configuration used for optimization, considerable spin squeezing is obtained.

Noise, changing at the time scale of a single measurement, can be simulated by shot-to-shot fluctuating rotation angles and interaction times acting on the bare resources. In Fig.~\ref{fig:OptExactExps} (bottom right panel) the optimal squeezer is affected by correlated noise, independent of system size $N$: the application time/angle of each bare global gate (see SM) suffers the same relative error, sampled from a normal distribution with standard deviation $\sigma_{\text{noise}}$. The data points are calculated by averaging over 10000 projective measurements each obtained from a wavefunction evolved under a different error realization. As expected, the impact of noise increases with the number of unitaries, thus a specific noise amplitude identifies an optimal depth $n$ of the squeezer.
Additionally we expect that a feedback-loop optimization in the presence of noise will improve these results, by adjusting the optimal solution according to the noise. 

To take into account the impact of spontaneous emission,
we compare the interaction timescale $V^{-1}_0$ to the rate of spontaneous emission due to the Rydberg dressing ($1/\tau_{\textrm{eff}}$), through the ratio $\eta_c = \tau_{\textrm{eff}}/V_0^{-1} = (\frac{\Omega_R}{2\Delta}) (\Omega_R \tau_R)$, where $\tau_R$ is the lifetime of the Rydberg state. For $\tau_R \sim 50 \mu\textrm{s}$, $\Delta = 10 \Omega_R$, and $\Omega_R = 2\pi\cdot20~\textrm{MHz}$, we expect $\eta_c \approx 300$.
Simulations show that the total timescales for optimal spin-squeezing are comparatively short: of the order of a few $V_0^{-1}$ for $R_C > a$  (see SM). In 2D they exhibit a sub-linear growth with depth $n$.
Moreover, running the quantum algorithm directly on the experimental platform automatically adjusts the optimal solution accordingly.

{\it Outlook --} 
Desirable properties of the optimized spin-squeezed states can be enforced by appropriately modifying the cost function (see SM).
Beyond that, quantum algorithms for metrology can be developed in a broader context, where the encoding, probing, decoding, and measurement steps are altogether variationally optimized. This will require efficient estimation techniques for general metrological cost functions, such as the Fisher information \cite{CavesFisherinformation}. The quantum algorithm we presented can be readily translated for different experimental architectures, involving e.g.~molecules or optical lattices, employing their respective programmable entanglement resource to generate spin-squeezing.

\begin{acknowledgments}
{\it Acknowledgements --}
We thank M. Norcia, A. Young, W. Eckner for discussions.
Work at Innsbruck is supported by the EU Quantum Technology Flagship project PASQUANS and the QuantEra project QTFLAG,
and the Austrian Research Promotion Agency (FFG) via QFTE project AutomatiQ.
Work at Boulder is supported by
the US Army Research Office grant W911NF-19-1-0223,
and the US Air Force Office of Scientific Research (AFOSR) grant FA9550-19-1-0275.
Numerical results were obtained via the LEO cluster of the Innsbruck University.
\end{acknowledgments}

\input{VQS-Merged.bbl}

\newpage

\section{Supplemental Material}

Below we collect the supplemental material for {\it Variational spin-squeezing on programmable quantum sensors}, which is organized as follows: In section I, we introduce the notion of the permutation invariant subspace and discuss the permutation symmetry-breaking caused by finite-range interactions. In section II, we discuss the implementation of the fundamental building blocks of the variational gate sequence and provide deeper insights into the design of the sequence with respect to efficiency requirements. Section III contains a detailed analysis of the optimal gate sequence obtained by the variational algorithm. In particular, we visualize the dynamics of the many body state by means of the Husimi distributions on the Bloch sphere.
 We highlight the most prominent differences between our variational quantum algorithm for metrology and variational quantum eigensolvers in section IV.

\section{Broken permutation symmetry: Many body dynamics with finite range interactions}

In this section, we discuss the breaking of permutation symmetry, as a particular aspect of variational state preparation with finite range interactions. We first introduce the notion of the permutation symmetric subspace in the context of one-axis twisting (OAT) and two-axis twisting (TAT). We then analyze mechanisms of variational preparation of spin-squeezed states, when finite range interactions break the permutation symmetry of the initial state.

\subsection{Permutation symmetric subspace}

As discussed in the main text, spin squeezed states can be prepared from a spin aligned product state $\ket{\uparrow_x}^{\otimes N} $ as quench dynamics generated by one-axis and two-axis twisting Hamiltonians, given by $H_{\rm OAT} \propto J_z^2$ and $H_{\rm TAT } \propto J_z^2 - J_y^2$, respectively.
These Hamiltonians correspond to a many-body systems with {\em infinite range} pairwise interactions, implying that they are invariant under particle permutation.  Let $\Gamma_{k \ell}$ be the operator that exchanges two particles $k$ and $\ell$. It follows that $[H_{\text{OAT}}, \Gamma_{k\ell}] = 0$ and $[H_{\text{TAT}}, \Gamma_{ij}] = 0$ for any p§air of particles $k$, $\ell$. Since the initial state is an eigenstate of the permutation operator: $\Gamma_{k\ell} \ket{\uparrow_x}^{\otimes N} = (+1) \ket{\uparrow_x}^{\otimes N}\ \forall k,\ell$, the many body dynamics after the quench will be constrained to the permutation-invariant subspace of linear dimension $(N+1)$. This subspace coincides with the eigenspace of the angular momentum operator with largest value $j=N/2$, so that $\boldsymbol{J}^2 |\Psi\rangle =  N/2(N/2+1)|\Psi\rangle$. Here we will use the collective angular momentum basis, or Dicke basis \cite{DickeBasis} $|j,m,\alpha\rangle$, where the index $\alpha$ labels  degeneracies: $\boldsymbol{J}^2 |j,m,\alpha\rangle =j(j+1)|j,m,\alpha\rangle$ and $J_z |j,m,\alpha\rangle= m|j,m,\alpha\rangle$. Each angular momentum `shell' $j$, i.e.~eigenspace of $\boldsymbol{J}^2$, is \[d_j^N=\frac{N!(2j+1)}{(N/2-j)!(N/2+j+1)!}\] fold degenerate. The permutation invariant subspace, or outer shell, spanned by $\{|N/2,m,1\rangle\}$ with ${-N/2\leq m \leq N/2}$, has no $\alpha$ degeneracy.
Notably, the {\em maximally} spin-squeezed state \cite{Brif1996, OptimalSqueezing}
lives within
this outer Bloch shell, and exhibits a spin-squeezing $\xi^2 = 2/(N+2)$. It can be prepared exactly with a time-dependent Hamiltonian dynamics using
$J_z^2$ and $J_x$ as resources
(i.e.~employing only $D_z$ and $R_x$ gates for $R_C = \infty$)
\cite{OptimalControlSqueezing}, as  shown in Fig.~\ref{fig:InfiniteRangedScaling}.

\begin{figure}[t]
\includegraphics[]{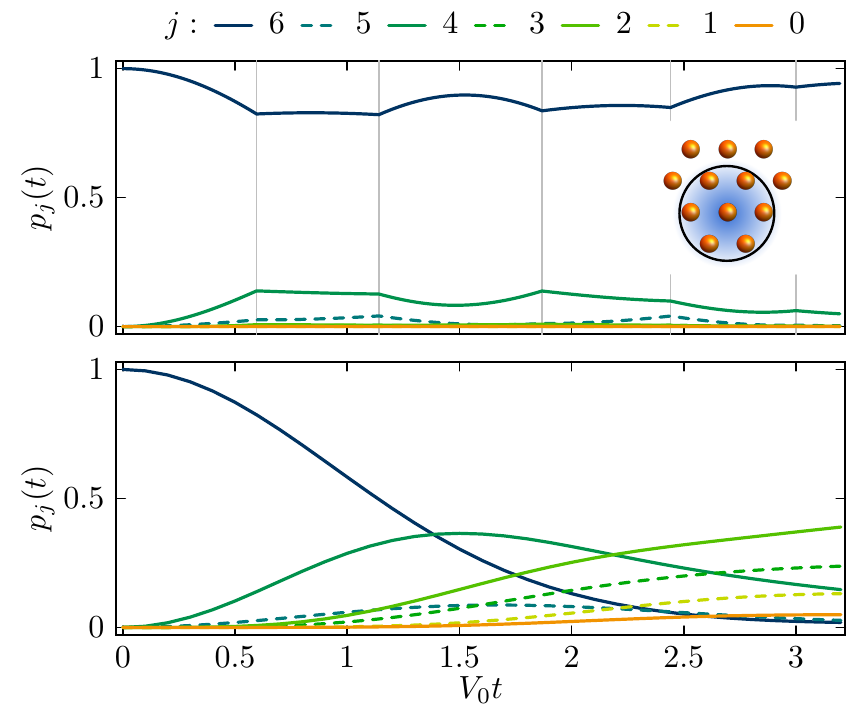}
\caption{Top: Populations $p_j$ of the many-body wavefunction, over the various eigenspaces $j$ of the total angular momentum $\boldsymbol{J}^2$, plotted in real-time for an optimal pulse sequence. The system is  a triangular lattice with $N=12$ atoms, interaction radius $R_c = 1.3$ and circuit depth $n = 3$. Bottom: Populations $p_j$ under evolution of a single $D_z$ gate, for the same system parameters.
\label{fig:ShellOccupation}}
\end{figure}

\subsection{Breaking of permutation symmetry with finite-range interactions}

Any pairwise interaction that is range-dependent, e.g.~Rydberg-dressing, provides entangling resources which break permutation invariance. Thus, even when starting from a symmetric initial state $\ket{\uparrow_x}^{\otimes N}$, the quench dynamics will populate lower angular momentum (Bloch) shells $J<N/2$ with time (see Fig.~\ref{fig:ShellOccupation}, bottom panel).

Numerical simulations show that our optimization algorithm attempts to variationally restore the permutation symmetry. This is demonstrated in Fig.~\eqref{fig:ShellOccupation} (top panel), where the populations 
\[p_j(t) = \sum_{m = -N/2}^{N/2}
\sum_{\alpha = 1}^{d_j^N}
|\langle j,m,\alpha | \Psi(t)\rangle|^2\]
of the wavefunction $|\Psi(t)\rangle$, over the various eigenspaces $j$ of $\boldsymbol{J}^2$, are plotted as a function of time $t$.

The optimized trajectories do explore transient stages where lower angular momentum shells become populated, but the population in the outermost shell $p_{j\ =\ N/2}$ never drops significantly below 1, and population is pushed back eventually towards $1$ at the end of the circuit.
Such behaviour emerges in all optimization runs we performed, regardless of the spatial geometry. In this figure, a triangular 2D lattice was considered with $R_C/a = 1.3$) and a circuit depth $n=3$. We conclude that the (partial) restoration of permutation invariance is a general feature of our variational algorithm, since the evolution of under a single time-independent $D_z$ (bottom panel) populate non-permutation invariant subspaces.

\begin{figure}[t]
\includegraphics[]{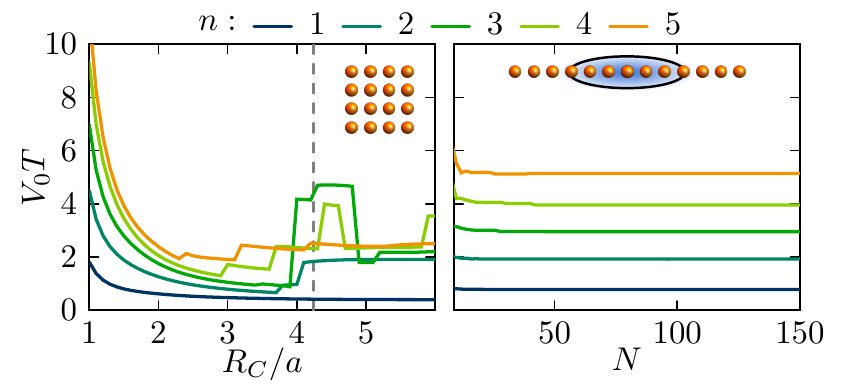}
\caption{Total interaction times $T = \sum_i (\tau_i + \tau'_i)$ of the optimal gate circuits for the exact optimization. Curves in two panels correspond to the optimal results in the two top panels of Fig.~3 (main text).
\label{fig:OptExactTotalnteractionTime}}
\end{figure}

\subsection{Total interaction time}
Since the variationl algorithm attempts to find trajectories that do not disrupt the particle permutation symmetry completely the resulting tend to grow slowly. In Fig.~\ref{fig:OptExactTotalnteractionTime} the total interaction times $T=\sum_{i=1}^{n} (\tau_i + \tau'_i)$ resulting from the exact optimization results (Fig.~3 main text) are displayed, in units of $V_0^{-1}$ ($\hbar = 1$). Since global rotations $R_a$ are implemented at timescales faster than $V_0^{-1}$, $T$ can be regarded as actual preparation time, and is the relevant quantity to compare when considering decoherence effects and errors occurring in the preparation stage.

The 2D case (left panel) reveals that, while nearest neighbour interaction radii $R_C\approx1$ yield large optimal $T$ due to the vanishing interaction amplitudes, intermediate and large ranges yield faster timescales $T$. Apart from a few numerical instabilities, deriving from the competition between compatible optima in the parameter landscape, all resulting preparation times in this regime are of the order of a few $V_0^{-1}$.

In contrast, the 1D case (right panel) offers a different perspective. While the optimal timescale is barely affected by system size, since boundary conditions become quickly negligible in $N$ and optimization of a translationally-invariant bulk becomes the main task, the preparation times exhibit a sub-linear increase in the circuit depth $n$, a result of the gradual buildup of entanglement, with $n$.

\section{Design of elementary gates and the variational gate sequence}
\subsection{Entangling gates from bare quantum resources}
As specified in the main text, the Rydberg dressing Hamiltonian $H_D$ and (global) single particle rotations specify the bare quantum resources of our model.  Below we discuss how $D_{x,z}$ gates are obtained from these bare resources. Following \cite{PohlSqueeze2014}, The site dependent single body light shifts $\sum_{i=1}^N \delta_i s_i^z$ appearing in the bare dressing Hamiltonian $H_D$ can be cancelled by using a spin-echo pulse,
\begin{align}
    D_z(\tau) &= R_{x}(\pi) \exp\left[ -i \frac{\tau}{2} H_D \right] R_{x}(\pi) \exp\left[ -i \frac{\tau}{2} H_D \right]\notag \\
    &= \exp\left[ -i \tau \sum_{i<j}^N V_{i,j} s_i^z s_j^z \right].
\end{align}
Here $R_y(\pi)$ can also be replaced by $R_x(\pi)$.
Additionally also static light shifts present will be cancelled by this spin-echo.

Interactions in other directions, different from $z$, the $D_z$, are obtained via rotations,
\begin{align}
    D_{x,y}(\tau)&=R_{y,x}(\pi/2) D_z(\tau)R_{y,x}(3\pi/2)\notag \\
    &= \exp\left[-i \tau \sum_{i<j}^N V_{i,j} s_i^{x,y} s_j^{x,y}\right].
\end{align}

\subsection{Preservation of the collective spin direction}

Here we show that the gate circuit of Eq.~(3) in the main text preserves the direction of the Bloch vector along the $x$ axis, $\langle \boldsymbol{J} \rangle_{\boldsymbol{\theta}}$, for an initial state $|\uparrow_x\rangle^{\otimes N}$.

Every gate in the sequence commutes with the parity operator in the $x$ direction $P_x=\prod_{i=1}^N s_i^x$, since
$P_x s^{a}_j P_x = (s^x s^a s^x)_j = \pm s^a_j$ and thus
$P_x D_{x,z} P_x = D_{x,z}$.
Since the initial state $\ket{\uparrow_x}^{\otimes N}$ is a $P_x$ eigenstate, the variational state $|\Psi(\boldsymbol{\theta})\rangle$ will remain an eigenstate, i.e.~$P_x |\Psi(\boldsymbol{\theta})\rangle = \pm|
\Psi(\boldsymbol{\theta})\rangle$.
This implies that $\langle J_y\rangle_{\boldsymbol{\theta}}=
\langle P_x J_{y,z} P_x \rangle_{\boldsymbol{\theta}}=
- \langle J_{y,z}\rangle_{\boldsymbol{\theta}}= 0$, which translates into a Bloch vector $\langle \boldsymbol{J} \rangle_{\boldsymbol{\theta}}$ directed by construction along the $x$ axis.

\subsection{Design of the variational circuit}

Below we motivate the design of the variational ansatz described in Eq.~(3) and Fig.~2 of the main text. This ansatz is employed in the variational quantum algorithm to generate spin-squeezing with Rydberg dressing resources. The basic building block is provided by the interaction gate $D_z$, which we combine with single-spin rotations. 

We restrict the spin-rotations, $R_{\boldsymbol{v}}(\theta) = \exp(-i \theta
\boldsymbol{v} \cdot \boldsymbol{J})$ to be global only but allow for arbitrary directions $v$, with $|\boldsymbol{v}| =1$). Consequently, the number of variational parameters in our ansatz will scale with the circuit depth $n$ only (and will be independent of the system size $N$).
As described before a  finite-range Ising gate
$D_{\boldsymbol{v}}(\tau) = \exp[-i \tau \sum_{i<j}^N V_{i,j} (\boldsymbol{s}_i \cdot \boldsymbol{v})
(\boldsymbol{s}_j \cdot \boldsymbol{v}) ]$ can be generated along any direction $\boldsymbol{v}$.

We now constrain the general gates  $R_{\boldsymbol{v}}$ and $D_{\boldsymbol{v}}$, to satisfy the preservation of the collective spin direction. Enforcing this property removes the need to determine the direction of the Bloch vector via measurements after the gate (sequence), which would require additional experimental runs.
This property is tied to the protection of the parity symmetry $P_x$. Restricting to gates commuting with $P_x$ reduces rotations to $R_x$, and interaction gates to $D_x$ and $D_{\boldsymbol{v}_\perp}$, where $\boldsymbol{v}_{\perp}$ is within the $y$-$z$ plane. Note here that any $D_{\boldsymbol{v}_\perp}$ can be generated from $R_x$ and $D_z$. The relevant gates fulfilling the requirement of axis preservation are, therefore, $R_z, D_x, D_z$.

Sequencing these three types of allowed gates, with equal repetition to generate the fastest possible growth of accessible states manifold, results in a sequence of layers of the form
$\bar{\mathcal{U}} = D_x(\tau') R_x(\varphi) D_{z}(\tau)$. 
We conclude that the sequence in Eq.~(3) (main text) comprises the most general set of gates satisfying all the aforementioned requirements.

\subsection{Complete gate sequence with bare quantum resources}
The sequence motivated above translates in to a series of $9n+1$ laser pulses. In particular each unitary in Eq.~(3) (main text) is decomposed into
\begin{align}
    \mathcal{U}_i=&R_y(\frac{\pi}{2})e^{-i H_D \frac{\tau'_i}{2}}R_y(\pi)e^{-i H_D \frac{\tau'_i}{2}}R_y(\frac{\pi}{2})\notag \\ &\times R_x(\vartheta_i)e^{-i H_D \frac{\tau_i}{2}}R_x(\pi)e^{-i H_D \frac{\tau_i}{2}}.
\end{align}

\section{Husimi distributions: visualizing an optimized sequence}
\begin{figure*}[t]
\includegraphics[width=\textwidth]{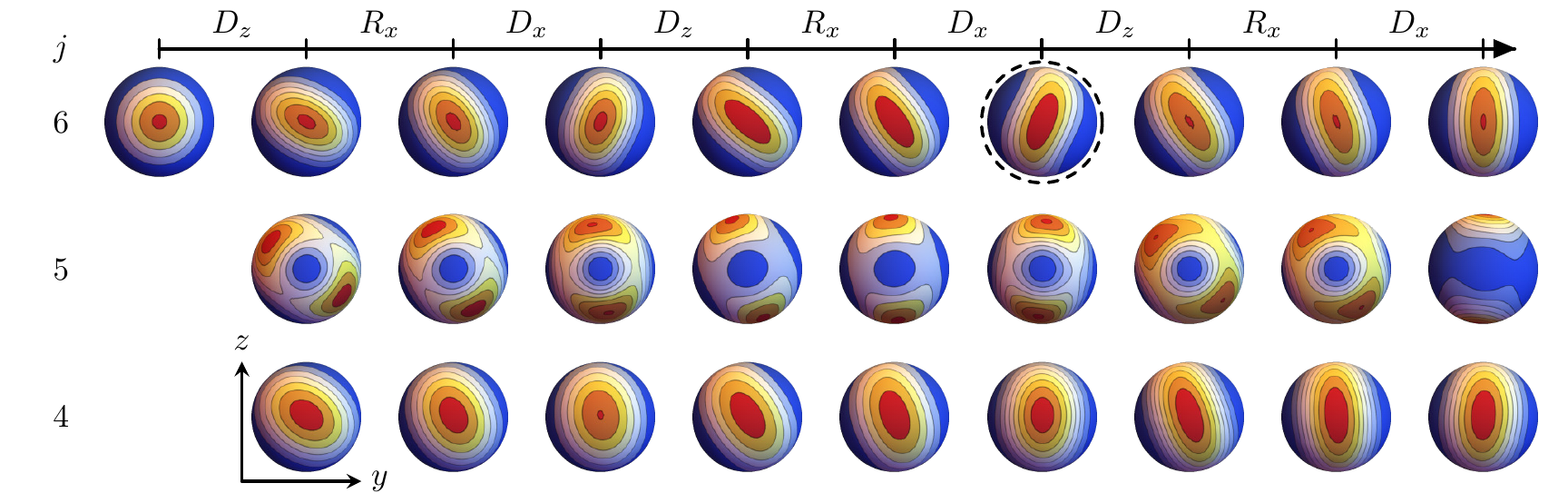}
\caption{Husimi-distributions of the three largest angular momentum shells, for the same optimized gate sequence shown in Fig.~\ref{fig:ShellOccupation}. The spheres are plotted after each gate in the sequence. The Husimi distribution marked with a dashed circle highlights a transient state which is visibly non-Gaussian.
\label{fig:HusimiFigure}}
\end{figure*}

Here we discuss the Husimi- distributions used in Fig.~2 of the main text in more detail and visualize their action in an optimized gate sequence. 

\subsection{Husimi distribution on the generalized Bloch sphere}
We employ the Husimi Q-representation to visualize the collective spin states on a generalized Bloch sphere. The Q distribution is defined as the overlap with a coherent spin-state
\[ \ket{\phi,\vartheta}=\sum_{m=-N/2}^{N/2}c^{N/2}_m(\vartheta)e^{-i(N/2+m)\phi}\ket{N/2,m,1},\] where \[c^j_m(\vartheta)=\binom{2j}{j+m}^{1/2}\cos(\vartheta/2)^{j-m}\sin(\vartheta/2)^{j+m}.\] 
This visualization is particularly relevant if the many body state is located in permutation invariant subspace. To be able to capture more information we generalize this to all total angular momentum eigenspaces, i.e.
\begin{align}
    Q_{\boldsymbol{\theta}}^j(\phi, \vartheta)=\Tr\left[\rho^j(\phi, \vartheta) \rho(\boldsymbol{\theta})\right]
\end{align}
where $\rho(\boldsymbol{\theta})=\ket{\Psi(\boldsymbol{\theta})}\bra{\Psi(\boldsymbol{\theta})}$ is the density matrix of a state $\ket{\Psi(\boldsymbol{\theta})}$ prepared by control parameters $\boldsymbol{\theta}$ and 
\begin{align}
    \rho^j(\phi,\vartheta)=\sum_{m, m'=-N/2}^{N/2} c^j_m(\vartheta) c^j_{m'}(\vartheta)e^{-i\phi(m-m')}\notag \\ \times\sum_{\alpha=1}^{d_j^N}\ket{j,m,\alpha}\bra{j,m',\alpha}.
\end{align}
If this quasi-probability distribution is plotted on a generalized Bloch sphere it visualizes the mean spin direction and the fluctuations of the collective spin around it, whereas fluctuations along the spin direction are not visible.

\subsection{Evolution of Husimi distributions in an optimized sequence}

Fig.~\ref{fig:HusimiFigure} displays, by means of the Husimi distributions, the quantum state after each gate of the optimized sequence shown in Fig.~\ref{fig:ShellOccupation}. We plot only the three largest shells $j = 4,5,6$ since their combined population remains above $99\%$ throughout the dynamics. The shell $j=4$ shows higher overall population than $j=5$ due to the fact that most permutation-symmetry breaking processes are double spin-flips, 
Interestingly, Husimi distributions of the even shells look similar, and both display spin-squeezing visibly, signalled by elliptic distributions. On the contrary, transient non-Gaussian states (see e.g.~ S-shaped distribution on the sphere indicated by a dashed circle) are visible only in the outer shell.

\subsection{Relevance of $D_x$ gate in a finite-range Sequence}
\begin{figure}[b!]
\includegraphics[]{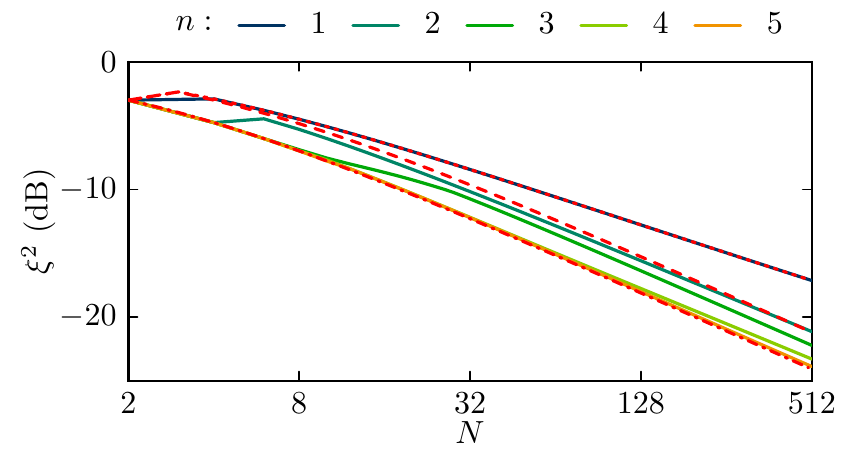}
\caption{Variational optimization of the spin-squeezing in the infinite range case, using a simplified variational circuit sequence that does not include $D_x$ (see text). The optimal squeezing is displayed as a function of the system size $N$ up to 512 (accessible due to the permutation symmetry), for various circuit depths $n = 1$ to 5. Results for OAT, TAT, and best achievable squeezing for $N$ particles, are drawn for reference as dashed lines.
\label{fig:InfiniteRangedScaling}}
\end{figure}
In this section, we discuss the role of the $D_x$ gate in sequence relying on finite-range interactions, in contrast to infinite range interactions $R_c \to \infty$, where it was shown that $R_x$ and $D_z$ are sufficient resources to generated spin-squeezing up to Heisenberg scaling $\xi^2 \propto N^{-1}$ \cite{LiuYouSqueeze2011,Shen2013,SqueezingTrotter2014,OptimalControlSqueezing}.
We recover these results in Fig.~\ref{fig:InfiniteRangedScaling}, where every layer of the variational circuit ansatz is simplified to
$\mathcal{U}'_i = R_x(\vartheta_i)D_z(\tau_i)$ and depends on 2 parameters instead of 3. Exact optimization results show that already for depth $n = 5$, spin-squeezing is saturated to its possible minimum $2/(N+2)$ up to 512 articles, i.e.~gaining a static prefactor (in $N$)  over TAT.
We motivate the discrepancy between the finite and infinite range scenarios with the following argument:
When $R_C = \infty$, under the action of $D_z(\tau)$, an initial coherent state will first evolve into a spin squeezed state followed by a non-Gaussian state, signaled by an S-shaped  Husimi-distribution \cite{RMPSqueezing}. In this case $D_z$ interaction is $4 \pi / V_0$-periodic, an intermediate S-shaped state can be transformed back into a Gaussian state. This property breaks down for finite-range Hamiltonian where $D_z$ is not periodic, and thus can not be inverted. Therefore a sequence consisting only of $R_x$ and $D_z$ is limited to operate at finite-range OAT times, because, at longer times, irreversible non-Gaussian states are generated. Numerical evidence shows that the $D_x$ gate is sufficient to revert the non-Gaussian state generation, thus lifting the short-time limitation which in turn limits the obtainable squeezing. We highlight this property in Fig.~\ref{fig:HusimiFigure}, where the Husimi distributions of the three largest momentum eigenspaces, are displayed after every gate of an optimized sequence. The sequence shows transient non-Gaussian states, marked by the dashed circle.

\section{Comparison to variational quantum eigensolvers}

The variational spin-squeezing algorithm we proposed fits within the framework of hybrid quantum algorithms, such as
Variational Quantum Eigensolvers (VQE) and simulators \cite{VQARahmani2013,Peruzzo2014,VQATroyer2015,VQAMcClean2016,Kandala2017,VQSError2017,MollGambetta2018,Kokail2019,Klco_2018,QAOAPichler2018},
and Quantum Approximate Optimization Algorithms (QAOA)
\cite{Farhi2014,Farhi2016,zhou2018quantum,QAOAPichler2018}. The goal of these algorithms is to variationally engineer on a quantum device ground states $|\Psi_0\rangle$ of target Hamiltonians $H_T$, either quantum or classical.
In this section, we highlight a few main conceptual differences between VQE/QAOA and the our spin-squeezing strategy.

First of all, while for VQE the cost function is the expectation value $\langle H_T \rangle_{\boldsymbol{\theta}}$ of the target Hamiltonian, thus infinitely differentiable and with bounded derivatives, cost functions for entanglement-enhanced quantum metrology problems are generally not simple expectation values. Known examples beyond the spin-squeezing parameter of Eq.~(1) (main text) include: the Fisher Information
\begin{equation}
F(O,t) = \sum_{k} \langle \Psi(t) | P_k | \Psi(t) \rangle\left(\frac{\partial}{\partial t} \log \langle \Psi(t) | P_k | \Psi(t) \rangle\right)^2,
\end{equation}
with $P_k$ eigenspace projectors of the observable $O$, quantifying the maximal sensitivity to $t$ of the projective measurements of $O$; the Quantum Fisher Information (see main text), which maximises $F(O,t)$ over all possible observables $O$ \cite{CavesFisherinformation,QFI_Entanglement_1, QFI_Entanglement_2}; and the Allan deviation, used in precision clocks to synchronize tunable optical resonators to the atomic frequency.
Adopting these quantifiers as cost functions for variational quantum algorithms requires not only the development of efficient estimation techniques for the cost function itself, but also the development of quantum algorithms not incurring in optimization instabilities (caused e.g. by singularities of the cost function, or its differentials, in the parameter landscape).

As a second fundamental difference, we remark that for VQE a single correct solution exists, and the degree of approximation is measured in terms e.g.~of fidelity. By contrast, for quantum metrology, {\it any} engineered state offering precision enhancement over SQL is a valid solution, and the challenge involves achieving the best possible enhancement for a limited set of available resources. This property allows additional freedom in the choice of cost function, as we discuss later on.

Finally, we stress that for optimizing spin-squeezing, the precision of the cost function estimator does not decrease with the system size $N$, for a fixed number of measurements. This property, which is not guaranteed for VQE, is discussed in detail later in this section.

\subsection{Modifying the cost function}
\begin{figure}[t]
\includegraphics[]{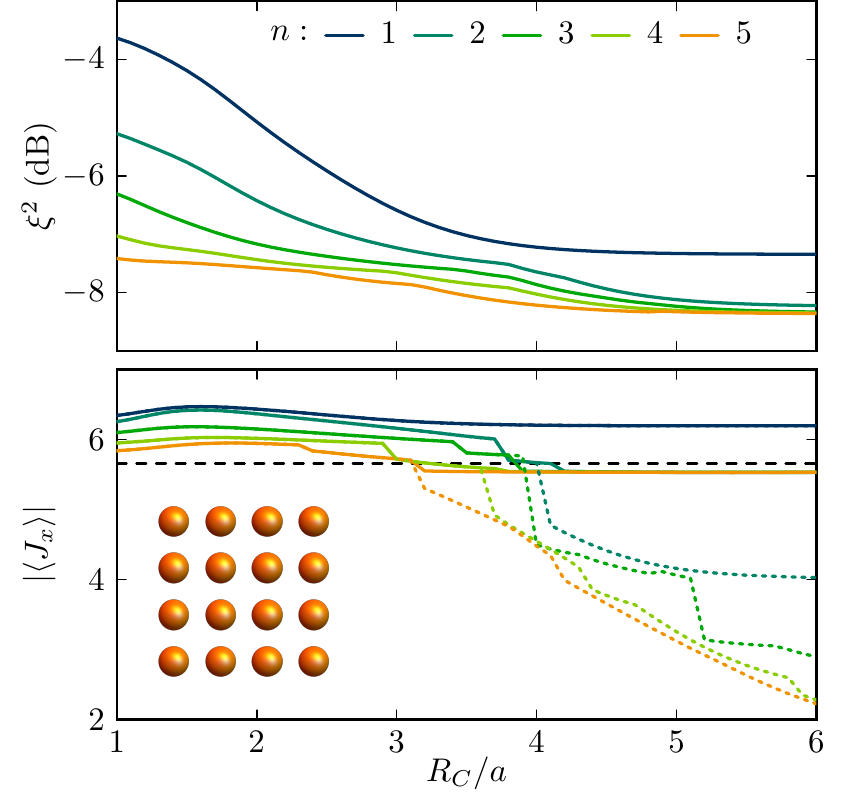}
\caption{Exact optimization results of the modified cost function in EQ.~\eqref{eq:modifiedCostfunction} for circuit depths ranging from $n =$ 1 to 5 layers  Top panel: squeezing parameter $\xi^2$ in a $4\times4$ square array as a function of the interaction radius $R_c/a$. Bottom panel: Bloch vector length corresponding to the same optimal solutions displayed in the top panel. The dotted lines show the spin length of the states in Fig.~3 in the main text, which have been optimized with the unmodified cost function. The horizontal dashed line indicates the threshold value $N/\sqrt{8}$. 
\label{fig:modifiedCostfunction}}
\end{figure}
As we discussed in the previous section, our goal is not to prepare a unique quantum many body state with high precision, but rather to enhance a collective property of the prepared wavefunction (in our case, the signal-to-noise ratio of Ramsey interferometry). This objective allows us some freedom in building the cost function itself, for instance, we can include additional {\it penalty} terms to favor other properties of the optimized state.
Penalties become particularly relevant for the problem of spin-squeezing, since the maximally spin-squeezed state for even $N$, the Dicke state $\lim_{\epsilon \rightarrow 0}\sqrt{1-\epsilon^2}\ket{N/2,0}+\epsilon/\sqrt{2}(\ket{N/2,+1}+\ket{N/2,-1})$, exhibits zero Ramsey signal due to its zero Bloch vector length $|\langle \boldsymbol{J} \rangle| = 0$. To avoid this pitfall, we can modify the cost function, ensuring a finite Bloch length by penalizing lengths below a certain threshold $\bar{x}$.


In particular we modify the cost function according to 
\begin{align}
    C(\boldsymbol{\theta})=N\frac{\braket{J^2_y}_{\boldsymbol{\theta}}}{\braket{J_x}^2_{\boldsymbol{\theta}}}+f_p(\left|\braket{J_x}_{\boldsymbol{\theta}}\right|)
    \label{eq:modifiedCostfunction},
\end{align}
where the penalty function $f_p(\cdot)$ is calculated on the same estimator $\braket{J_x}_{\boldsymbol{\theta}}$ that we employ for $\xi^2$, ensuring that we can still run the feedback-loop on the experiment, at the same cost.
Fig.~\ref{fig:modifiedCostfunction} shows numerical simulations where we adopted the penalty function
\begin{align}
f_p(x) = 
    \begin{cases}
        e^{1/\left(x - \bar{x}\right)} & 0 < x < \bar{x} \\
        0 &  \bar{x} \leq x
    \end{cases},
\end{align}
which is monotonic and $\mathcal{C}^{\infty}$ on its domain $\mathbb{R}^+$ to guarantee smooth convergence of the search algorithm.
We set a threshold $\bar{x} =N/\sqrt{8}$, which is known to allow a Heisenberg like scaling \cite{OptimalControlSqueezing}.
The top panel shows optimized spin squeezing according to the modified cost function in a scenario analogous to Fig.~3 in the main text. Here we observe that the short range regime ($R_C<3a$) remains unchanged. By contrast, for large interaction radii the optimal squeezing saturated earlier in the depth $n$. The long range solutions in Fig.~3 in the main text approach the maximally spin squeezed state with vanishing Bloch vector length, indicated by the dotted lines in the lower panel. 

\subsection{Measurement scaling with system size}

\begin{figure}[t]
\includegraphics[]{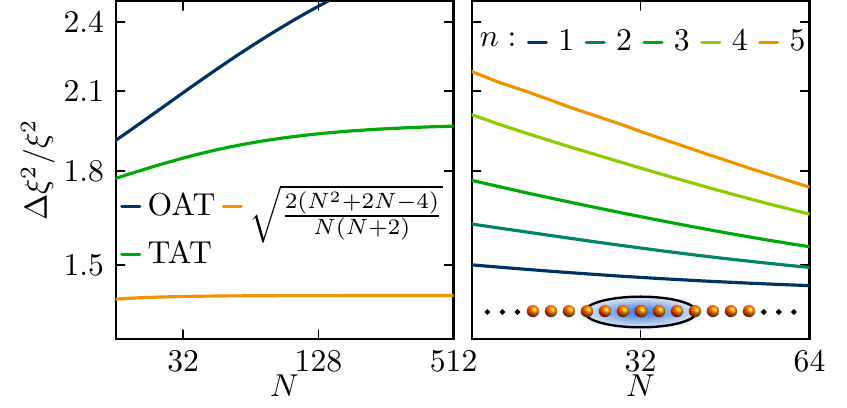}
\caption{Left panel: relative squeezing error for the optimal states generated by OAT, TAT and a trial state, discussed in the text. Right panel: Relative error for the optimized spin squeezed states in a 1D Array at $R_C/a=3$ (see Fig.~3 main text for corresponding squeezing values). 
\label{fig:measuremetnscaling}}
\end{figure}

In this section, we investigate the number of measurements required to determine the spin-squeezing parameter to a given relative precision $\left(\Delta \xi^2\right)/ \xi^2 $, as a function of the system size. Our analysis reveals that for variationally optimized states, the relative error of the spin squeezing parameter saturates for large particle numbers. 
This  is in contrast to variational quantum eigensolvers, where, for example, the extensiveness of the expectation value of a quantum-chemistry Hamiltonian, requires an increasing amount of measurements, as the number of particles increases.

We start by analytically calculating the standard deviation of the spin-squeezing parameter for optimally spin-squeezed states. To this end, we construct a permutation invariant state which exhibits spin-squeezing up to the Heisenberg-limit:
\begin{align}
    \ket{\Psi}=\frac{1}{\sqrt{2}}\ket{N/2,0}+\frac{1}{2}\left(\ket{N/2,+1}+\ket{N/2,-1}\right),
\end{align} in the Dicke basis ($\alpha=1$). 
We can calculate the squeezing of this state via the angular-momentum operators,
\begin{align} \label{eq:JzDicke}
J_z&=\sum_{M=-N/2}^{N/2} M \ket{N/2, M}
\end{align}
\begin{align}
\begin{split}
J_+=\sum_{M=-N/2}^{N/2-1} &\sqrt{(N/2+M+1)(N/2-M)}\\ &\times \ket{N/2,M+1}\bra{N/2,M}
\end{split}
\end{align}
\begin{align}
\begin{split}
J_-=\sum_{M=-N/2}^{N/2-1}& \sqrt{(N/2-M+1)(N/2+M)}\\ &\times \ket{N/2,M-1}\bra{N/2,M}
\end{split}
\end{align}
\begin{align} \label{eq:JxJyDicke}
J_x=\frac{1}{2}(J_++J_-), \ \ J_y=i\frac{1}{2}(J_+-J_-).
\end{align}
Using Eq.~(\ref{eq:JzDicke}-\ref{eq:JxJyDicke}), the spin-squeezing evaluates to
\begin{align}
	\xi^2=N\frac{\braket{J_z^2}}{\braket{J_x}^2}=\frac{4}{ (N+2)}.
\end{align}
In the following, we use Gaussian error propagation to determine the variance of $\xi^2$:
\begin{align}
\left(\Delta \xi^2\right)^2 = \left( \frac{\partial \xi^2}{\partial \braket{J_z^2}} \right)^2 \Delta J_z^2 + \left( \frac{\partial \xi^2}{\partial \braket{J_x}} \right)^2 \Delta J_x^2.
\end{align}
As a result, the relative squared error evaluates to 
\begin{align} \label{eq:relana}
	\frac{\left(\Delta \xi^2\right)^2}{\left( \xi^2 \right)^2} = \frac{2(N^2 + 2N - 4)}{N(N+2)} = 2 - \frac{8}{N^2} + \mathcal{O}\left(\frac{1}{N^3}\right).
\end{align}

Eq.~\ref{eq:relana} shows that the relative error saturates for large particle numbers $N$, thus the number of measurements to determine the spin-squeezing to a given relative precision approaches a constant. In Fig.~\ref{fig:measuremetnscaling} (left panel) we plot the relative error as a function of the system size $N$. The results for OAT and TAT have been obtained numerically, by evolving the initial state $\ket{\uparrow_x}$ with the OAT- and TAT-Hamiltonian, until the spin squeezing reaches its optimal value.

In Fig.~\ref{fig:measuremetnscaling} (right panel), we provide a numerical analysis of the relative error for a 1D chain of particles up to 64 sites, using matrix product states (MPS). The simulations are performed for $R_C/a=3$ and up to 5 layers for the variational circuit. We observe a decreasing relative error with $N$, which we attribute to the fact, that the the squeezing-parameter decreases sub-polynomial in $N$ in 1D see Fig. 3 upper right panel in the main text), different than the absolute error $\Delta \xi^2$ which decreases nonetheless polynomially.  We conclude that the optimization algorithm as a whole will not require an increasing amount of measurements as the system size is increased.

\end{document}

%% file: VQS-Merged.bbl
%